\documentclass[submission]{eptcs}

\providecommand{\event}{V2CPS 2016}
\title{%
    A Compositional Framework for Preference-Aware Agents
    \thanks{The work was partially supported by ONR grant N00014--15--1--2202.}
}
\newcommand{\titlerunning}{A Compositional Framework for Preference-Aware Agents}
\newcommand{\authorrunning}{T. Kapp\'e, F. Arbab \& C. Talcott}

\author{Tobias Kapp\'e
\institute{LIACS, Leiden University \\ Leiden, The Netherlands}
\institute{Centrum Wiskunde \& Informatica \\ Amsterdam, The Netherlands}
\email{tkappe@liacs.nl}
\and
Farhad Arbab
\institute{Centrum Wiskunde \& Informatica \\ Amsterdam, The Netherlands}
\institute{LIACS, Leiden University\\ Leiden, The Netherlands}
\email{farhad@cwi.nl}
\and
Carolyn Talcott
\institute{SRI International \\ Menlo Park, CA 94025, USA}
\email{carolyn.talcott@sri.com}
}

\usepackage[T1]{fontenc}

\usepackage{hyperref}

\usepackage{tikz}
\usetikzlibrary{positioning,arrows}
\usetikzlibrary{decorations,decorations.markings}

\usepackage{amssymb}
\usepackage{amsmath}
\usepackage{amsthm}

\newtheorem{definition}{Definition}

\usepackage{microtype}

\renewcommand{\restriction}{\mathord{\upharpoonright}}

\newcommand{\angl}[1]{\ensuremath{\left\langle #1 \right\rangle}}

\newcommand{\boldzero}{\mathbf{0}}
\newcommand{\boldone}{\mathbf{1}}

\newcommand{\SDC}{\mathsf{SDC}}

\newcommand{\SCSP}{\mathsf{SCSP}}
\newcommand{\Assign}{\mathsf{Assign}}
\newcommand{\Sol}{\mathsf{Sol}}

\newcommand{\tail}{\mathsf{tail}}
\newcommand{\cancel}{\mathcal{C}}
\newcommand{\proj}{\mathsf{Pr}}

\newcommand{\mleft}{\mathsf{left}}
\newcommand{\mright}{\mathsf{right}}
\newcommand{\mforward}{\mathsf{forward}}
\newcommand{\mbackward}{\mathsf{backward}}
\newcommand{\mstay}{\mathsf{stay}}
\newcommand{\turn}{\mathsf{turn}}
\newcommand{\charge}{\mathsf{charge}}
\newcommand{\discharge}{\mathsf{discharge}}
\newcommand{\rest}{\mathsf{rest}}
\newcommand{\energy}{\mathsf{energy}}
\newcommand{\progress}{\mathsf{progress}}
\newcommand{\backtrack}{\mathsf{backtrack}}
\newcommand{\converge}{\mathsf{converge}}
\newcommand{\diverge}{\mathsf{diverge}}

\pgfmathsetmacro{\gap}{3}
\pgfmathsetmacro{\angle}{15}
\pgfmathsetmacro{\nodesize}{10mm}
\pgfmathsetmacro{\scale}{0.8}

\usepackage{enumitem}

\def\since{3}

\newcommand{\change}[2]{%
\ifnum#1>\since%
    \textcolor{red}{#2}%
\else%
    #2%
\fi
}

\newcommand{\marginchange}[2]{%
\ifnum#1>\since%
    \marginpar{\tiny \textcolor{red}{#2}}
\fi
}

\usepackage{mathpartir}

\usepackage{subcaption}

\usepackage{framed}
\definecolor{shadecolor}{HTML}{eeeeee}

\usepackage{pdfpages}

\usepackage{booktabs}

\allowdisplaybreaks{}

\begin{document}
\maketitle

\begin{abstract}
A formal description of a Cyber-Physical system should include a rigorous specification of the computational and physical components involved, as well as their interaction. Such a description, thus, lends itself to a compositional model where every module in the model specifies the behavior of a (computational or physical) component or the interaction between different components. We propose a framework based on Soft Constraint Automata that facilitates the component-wise description of such systems and includes the tools necessary to compose subsystems in a meaningful way, to yield a description of the entire system. Most importantly, Soft Constraint Automata allow the description and composition of components' preferences as well as environmental constraints in a uniform fashion. We illustrate the utility of our framework using a detailed description of a patrolling robot, while highlighting methods of composition as well as possible techniques to employ them.
\end{abstract}

\section{Introduction}


As the complexity of tasks for Cyber-Physical Systems grows, so does the need for the ability to decompose the specification of a system into multiple components. Such a decomposition generally eases the design of the specification and furthermore improves reusability. Formal verification tools, like model checkers such as Vereofy~\cite{baier-blechmann-klein-kluppelholz-leister-2010}, for instance, can then separately verify the properties of individual components, simplifying verification of properties claimed to emerge from their composition.

To make such descriptions robust against obstructions, it makes sense to enrich them with a notion of preference~\cite{talcott-arbab-yadav-2015}. For instance, if a patrolling robot finds its path obstructed, it may settle for a slightly adjusted patrol path. More generally, the presence of lower-preference actions enables a component (and the system as a whole) to be more flexible when its most-preferred action is incompatible with the actions allowed by other components or, especially, its (less predictable) environment. Endowment of preferences and ``alternative'' behavior to components raises the question of how to go about composing them in a meaningful way. In particular, not all concerns may be equally important in a composition.

The contribution of this paper is a framework that enables component-wise, preference-aware descriptions of Cyber-Physical Systems, based on Soft Constraint Automata (SCAs)~\cite{arbab-santini-2013}, a generalisation of Constraint Automata~\cite{baier-sirjani-arbab-rutten-2006}. Because SCAs express preferences using well-studied structures called c-semirings~\cite{bistarelli-montanari-rossi-1995}, we can rely on existing results to develop methods to compose SCAs.

Our notion of preference differs from earlier investigations of priority in the context of concurrency theory~\cite{cleaveland-hennessy-1990}.  First, priority is usually assigned to favor specifically prioritized actions over non-prioritized ones in otherwise nondeterministic choices. Preferences represent more abstract constraints that express considerations and compromises among a broad spectrum of concerns that often cannot even be directly related with each other. Moreover, priority assignments usually induce a statically determined partial order on concrete actions. As more abstract constraints, preferences of an agent need not specify concrete actions, and compose with other concerns and constraints that arise from a dynamically changing environment to select a suitable action for the agent. We believe both of these factors contribute towards a more compositional framework.


The remainder of this paper is organised as follows. In Section~\ref{sec:preliminaries} we review the necessary notation. In Section~\ref{sec:components}, we describe the components of our framework, after which we expand on the methods for their composition in Section~\ref{sec:composition}. We give a detailed example of modelling a patrolling robot using the proposed framework in Section~\ref{sec:example}. We list our conclusions in Section~\ref{sec:conclusion} and discuss further work in Section~\ref{sec:further-work}.

\section{Preliminaries}
\label{sec:preliminaries}


To introduce the components of our framework, we first discuss some necessary notions. Most importantly, we review Constraint Semirings and Soft Constraint Satisfaction Problems later in this section.

Let $S$ be a set. When $S_1$ and $S_2$ are sets such that $S \subseteq S_1 \times S_2$, we write $\proj_i(S)$, $i \in \{1,2\}$, for the \emph{projection} to the $i$-th component of $S$, i.e., $\proj_i(S) = \{ s_i : \angl{s_1, s_2} \in S \}$. Also, we use $S^\omega$ to denote the set of \emph{streams}~\cite{rutten-2005} of elements in $S$, i.e., the set of functions from $\mathbb{N}$ to $S$. For $\sigma \in S^\omega$, we write $\tail(\sigma)$ for the unique element of $S^\omega$ such that $\tail(\sigma)(n) = \sigma(n+1)$. We may abbreviate the repeated application of $\tail$ by $\tail^{n}$; more precisely, $\tail^0$ is the identity function, while $\tail^{n+1} = \tail \circ \tail^n$ for $n \geq 0$.

We often need to work with sets of symbols that may have associated values from some domain $\mathbb{D}$. Throughout this paper, we consider $\mathbb{D}$ to be fixed. Let $V$ be a finite set (of symbols). An \emph{assignment} of $V$ is a function $\alpha: V \to \mathbb{D}$. We denote the set of all assignments of $V$ by $\Assign(V)$, while $\Assign_\subseteq(V)$ denotes the set of all assignments of subsets of $V$, i.e., $\Assign_\subseteq(V) = \bigcup \{ \Assign(V') : V' \subseteq V \}$.

If $\alpha$ is an assignment of $V$ and $V' \subseteq V$, then the restriction of $\alpha$ to $V'$, written $\alpha\restriction_{V'}$, is the unique assignment of $V'$ that coincides with $\alpha$ on $V'$. If $\alpha$ and $\beta$ are assignments of $V$ and $U$ respectively and both agree on $U \cap V$, then their composition, written $\alpha + \beta$, is the unique assignment of $U \cup V$ such that $(\alpha + \beta)\restriction_V = \alpha$ and $(\alpha + \beta)\restriction_U = \beta$.

\subsection{Preferences}
\label{sec:preferences}

In order to work with preferences, we need a domain for expressing them. Additionally, we require an operator to select the best preference value from a set of preference values (if such a best value exists) and an operator to compose preference values. It turns out that a Constraint Semiring provides the right kind of structure for such a domain. The exact definition below is due to Bistarelli~\cite{bistarelli-2004}. Here, the operator $\bigoplus$ models the selection of the best preference, while $\otimes$ is used to obtain the preference of a composed action, given the preferences of the component actions.
\begin{definition}[c-semiring]
\label{def:c-semiring}
A \emph{Constraint Semiring} (or \emph{c-semiring}, for short) is a tuple $\angl{E, \bigoplus, \otimes, \boldzero, \boldone}$ where $E$ is a set (called the \emph{carrier} of the semiring) with $\boldzero, \boldone \in E$, while $\bigoplus: \mathcal{P}(E) \to E$ and $\otimes: E \times E \to E$ are operators such that for every $e \in E$ and every family of subsets of $E$ indexed by some set $I$, ${\{E_i\}}_{i \in I}$:
\begin{itemize}
    \setlength{\itemsep}{0em}
    \item $\bigoplus$ obeys the following restrictions:
    \begin{itemize}
        \item $\bigoplus \{ e \} = e$
        \item $\bigoplus \emptyset = \boldzero$ and $\bigoplus E = \boldone$
        \item $\bigoplus \left( \bigcup_{i \in I} E_i \right) = \bigoplus \{ \bigoplus_{i \in I} E_i : i \in I \}$ (the \emph{flattening} property).
    \end{itemize}
    \item $\otimes$ is commutative and associative, with $\boldone$ its unit element and $\boldzero$ its absorbing element.
    \item $\bigoplus$ distributes over $\otimes$, i.e., $e \otimes \bigoplus E_i = \bigoplus \left\{ e \otimes e' : e' \in E_i \right\}$
\end{itemize}
\end{definition}

Examples of c-semirings include the \emph{boolean semiring} $\mathbb{B} = \angl{\{\bot, \top\}, \bigvee, \wedge, \bot, \top}$ the \emph{weighted (or tropical) semiring} $\mathbb{W} = \angl{\mathbb{R}_{\geq 0} \cup \{ \infty \}, \inf, \hat{+}, \infty, 0}$ and the \emph{probabilistic semiring} $\mathbb{P} = \angl{[0,1], \sup, \cdot, 0, 1}$. Also interesting is the \emph{set semiring} $\mathbb{S}_{\Sigma} = \angl{\mathcal{P}(\Sigma), \bigcup, \cap, \emptyset, \Sigma}$ where $\Sigma$ is some finite set of symbols.

We refer to the set $E$ as a c-semiring if the other elements are not relevant for the discussion. When we want to explicitly mention a single operator or unit of a semiring $E$, we write $\bigoplus_E$, $\boldzero_E$, and so on. Applying $\bigoplus$ on two-element sets, we denote it as a binary infix operator $\oplus$. A c-semiring $\angl{E, \bigoplus, \otimes, \boldzero, \boldone}$ induces a relation $\leq_E$ on $E$, defined for $e, e' \in E$ as $e \leq_E e'$ if and only if $e \oplus e' = e'$. The relation $<_E$ on $E$ is $\leq_E$ with its diagonal excluded. In general, $\leq_E$ is a partial order~\cite{bistarelli-2004}; one can construct c-semirings (such as $\mathbb{S}_A$ when $|A| > 1$) where the induced ordering is not total. Constraint Semirings have a number of convenient properties. We refer to~\cite[Section 2.1]{bistarelli-2004} for an excellent treatment including proofs.

Particularly noteworthy in a c-semiring are cancellative elements:
\begin{definition}[cancellative elements]
\label{def:cancellative-element}
Let $\angl{E, \bigoplus, \otimes, \boldzero, \boldone}$ be a c-semiring. An element $e \in E$ is \emph{cancellative} if for all $e', e''$ we have $e' \otimes e \ =\  e'' \otimes e$ if and only if $e' = e''$. The set of cancellative elements of $E$ is written $\cancel(E)$, and its complement within $E$ is denoted by $\overline{\cancel}(E)$. When $\cancel(E) \cup \{ \boldzero \} = E$, we call $E$ a \emph{cancellative c-semiring.}
\end{definition}

Of course, one can construct new c-semirings from existing c-semirings $E_1$ and $E_2$. For instance, we can impose a c-semiring structure on the Cartesian product of two c-semirings, gaining the product semiring $E_1 \times E_2$~\cite{bistarelli-montanari-rossi-1997}. One can also trim down this product, by filtering out pairs containing $\boldzero_i$:
\begin{definition}[join semiring]
\label{def:join-semiring}
Let $\angl{E_1, \bigoplus_1, \otimes_1, \boldzero_1, \boldone_1}$ and $\angl{E_2, \bigoplus_2, \otimes_2, \boldzero_2, \boldone_2}$ be two cancellative c-semirings. Their \emph{join semiring}, written $E_1 \odot E_2$, is the c-semiring $\angl{E_1 \odot E_2, \bigoplus, \otimes, \angl{\boldzero_1, \boldzero_2}, \angl{\boldone_1, \boldone_2}}$ in which $E_1 \odot E_2$ is the set $(\cancel(E_1) \times \cancel(E_2)) \cup \{ \angl{\boldzero_1, \boldzero_2} \}$ and $\bigoplus$ and $\otimes$ are defined for $E' \subseteq E_1 \odot E_2$ and $e_i, e_i' \in E_i$ as follows:
$$
\bigoplus E' = \angl{\bigoplus\nolimits_1 \proj_1(E'), \bigoplus\nolimits_2 \proj_2(E') }
\quad \mbox{and} \quad
\angl{e_1, e_2} \otimes \angl{e_1', e_2'} = \angl{e_1 \otimes_1 e_1', e_2 \otimes_2 e_2'}
$$
\end{definition}
One can easily surmise that $E_1 \odot E_2$ is cancellative, and that the order induced on the join semiring $E_1 \odot E_2$ is the product order obtained from the orders induced on its operands, i.e., if $e_i, e_i' \in E_i$ then
$$\angl{e_1, e_2} \leq_{E_1 \odot E_2} \angl{e_1', e_2'} \iff e_1 \leq_{E_1} e_2\ \mathrm{and}\ e_1' \leq_{E_2} e_2'$$
The join semiring can be used to compose preferences from two c-semirings that are deemed equally important, or at least, a selection between the two indications of preference is yet to be made. In Section~\ref{sec:example}, we use it for our patrolling robot, where we want to defer selection of the most important preference (energy level versus the patrolling mission) to run-time.

Because we can construct a c-semiring that induces the product ordering on the Cartesian product, one may wonder if we can also construct a c-semiring that induces a lexicographic ordering. Indeed, this is possible, but to maintain distributivity of $\otimes$ over $\bigoplus$ we must exclude some elements from the carrier~\cite{gadducci-holzl-monreale-wirsing-2013}. As a consequence, not all types of c-semirings will be equally suitable as candidates to serve as the most significant component of a lexicographic composition.

\begin{definition}[lexicographic product semiring]
\label{def:lexicographic-product-semiring}
Let $\angl{E_1, \bigoplus_1, \otimes_1, \boldzero_1, \boldone_1}$ and $\angl{E_2, \bigoplus_2, \otimes_{2}, \boldzero_2, \boldone_2}$ be two c-semirings%
\footnote{In~\cite{gadducci-holzl-monreale-wirsing-2013}, $E_1$ is restricted to have a total ordering. We claim that such a restriction is not necessary for this construction to yield a c-semiring. A proof appears in the appendix of the technical report~\cite{techreport}.
}.
Their \emph{lexicographic product semiring}, written $E_1 \triangleright E_2$, is the c-semiring $\angl{E, \bigoplus, \otimes, \boldzero, \boldone}$ in which $E$ is the set $(\cancel(E_1) \times E_2) \cup (\overline{\cancel}({E_2}) \times \{ \boldzero_2 \})$, while $\boldzero = \angl{\boldzero_1, \boldzero_2}$ (and likewise for $\boldone$). Lastly, $\bigoplus$ and $\otimes$ are defined for $E' \subseteq E$ and $e_i,e_i' \in E_i$ by
$$
\bigoplus E' = \angl{\bigoplus\nolimits_1 \proj_1(E'), \bigoplus\nolimits_2 m(E') }
\quad \mbox{and} \quad
\angl{e_1, e_2} \otimes \angl{e_1', e_2'} = \angl{e_1 \otimes_1 e_1', e_2 \otimes_2 e_2'}
$$
in which $m(E')$ contains precisely the elements $e_2$ of $\proj_2(E')$ such that $\angl{\bigoplus\nolimits_1 \proj_1(E'), e_2} \in E'$.
\end{definition}
The lexicographic product semiring is useful when one concern is objectively more important than another. In Section~\ref{sec:example} we use it to mark the preference of patrolling as most important, while the preference to stay on the path is considered of secondary importance.

The intuition behind the second component of the additive operator above is that the best preference is chosen from the elements in the least significant position that co-occur with the best preference value in the most significant position --- since any value for which this is not the case occurs in a tuple that cannot be the maximal element of $E$ with respect to the lexicographic ordering.

To move between different domains of preferences in a smooth manner, the notion of a c-semiring homomorphism is useful. Its definition simply follows the familiar pattern from algebra. We will use homomorphisms later on to embed preference values from individual components into composed c-semirings.

\begin{definition}[c-semiring homomorphism]
Let $E_1$ and $E_2$ be c-semirings. A \emph{c-semiring homomorphism} (or in this paper, simply a \emph{homomorphism}) is a function $h: E_1 \to E_2$ such that
\begin{itemize}
    \setlength{\itemsep}{0em}
    \item $h(\boldzero_{E_1}) = \boldzero_{E_2}$ and $h(\boldone_{E_1}) = \boldone_{E_2}$
    \item for all $E_1' \subseteq E_1$ it holds that $h\left(\bigoplus_{E_1} E_1' \right) = \bigoplus_{E_2} \{ h(e) : e \in E_1' \}$
    \item for all $e, e' \in E_1$ it holds that $h(e \otimes_{E_1} e') = h(e) \otimes_{E_2} h(e')$
\end{itemize}
We call $h$ \emph{order-reflecting}~\cite{li-ying-2008} if for all $e, e' \in E$, it holds that $h(e) \leq_{E_2} h(e')$ implies $e \leq_{E_1} e'$.
\end{definition}
It can easily be shown that if $h: E_1 \to E_2$ is an order-reflecting homomorphism, then for all $e, e' \in E_1$ it holds that $e \leq_{E_1} e'$ if and only if $h(e) \leq_{E_2} h(e')$.

Let $E_1$ and $E_2$ be c-semirings. The following mappings are particularly useful:
\begin{align*}
h_L(e) =
\begin{cases}
\angl{\boldzero_{E_1}, \boldzero_{E_2}} & e = \boldzero_{E_1} \\
\angl{e, \boldone_{E_2}} & \mathrm{otherwise}
\end{cases}
&&
h_R(e) =
\begin{cases}
\angl{\boldzero_{E_1}, \boldzero_{E_2}} & e = \boldzero_{E_2} \\
\angl{\boldone_{E_1}, e} & \mathrm{otherwise}
\end{cases}
\end{align*}
If $E_1$ (respectively $E_2)$ is cancellative, then one can observe that $h_L$ (respectively $h_R$) is an order-reflecting homomorphism from $E_1$ (respectively $E_2$) to $E_1 \odot E_2$ as well as $E_1 \triangleright E_2$. In the sequel, we refer to these mappings as \emph{canonical injections}. Their domain and codomain c-semirings will always be made explicit. Lastly, the injection from $E_1 \odot E_2$ into $E_1 \times E_2$ is also an order-reflecting homomorphism.

\subsection{Soft Constraint Satisfaction Problems}
\label{sec:scsps}

The notion of a Soft Constraint Satisfaction Problem elegantly captures~\cite{bistarelli-montanari-rossi-1995} a number of generalizations of Constraint Satisfaction Problems aimed at attaching preference values to candidate solutions. Our definitions below are compatible with those of~\cite{bistarelli-montanari-rossi-1995,bistarelli-2004} but differ slightly for the sake of subsequent representation.

\begin{definition}[Soft Data Constraint]
A \emph{Soft Data Constraint} (SDC) is a tuple $\angl{U, E, [\cdot]}$ such that $U$ is a finite set of symbols, $E$ is a c-semiring (called the \emph{underlying semiring}) and $[\cdot]$ is a function from $\Assign(V)$ to $E$. We write $\SDC(V, E)$ for the set of all SDCs that involve a subset of $V$ and have $E$ as their underlying semiring.
\end{definition}

When $\angl{U, E, [\cdot]}$ is an SDC such that $[\cdot]$ has a range of $\{\boldzero, e\} \subseteq E$ with $e \neq \boldzero_E$, we call $\angl{U, E, [\cdot]}$ a \emph{binary constraint}. It is often more convenient to denote binary constraints in an abbreviated fashion. We write such constraints as $\angl{U, \phi, e}$, where $\phi$ is some first-order logic expression with $U$ as variables that is satisfied by $\alpha \in \Assign(U)$ if and only if $[\alpha] = e$. The c-semiring $E$ will always be clear from the context when we use this abbreviation. When $E = \mathbb{B}$, we omit $e$, for necessarily $e = \boldone_\mathbb{B} = \top$.

In a Soft Constraint Satisfaction Problem, one can use Soft Data Constraints to express preferences for assignments of subsets of variables. The total preference is then given by composing the preference values obtained from the SDCs into one.

\begin{definition}[Soft Constraint Satisfaction Problem]
A \emph{Soft Constraint Satisfaction Problem} (SCSP) is a tuple $\angl{V, E, C}$ such that $V$ is a finite set of symbols, $E$ is a c-semiring (called the \emph{underlying semiring}) and $C$ is a finite subset of $\SDC(V, E)$. We write $\SCSP(V, E)$ for the set of all SCSPs that involve a subset of $V$ and have $E$ as their underlying semiring.
\end{definition}
As an example of an SCSP, let $\mathbb{D} = \mathbb{R}_{\geq 0}$ and consider $K = \angl{\{x, y\}, \mathbb{P}, C}$ with $C$ containing precisely the (abbreviated) SDCs $\angl{ \{x \}, x < 1, 0.9 }$ and $\angl{ \{x,y\}, x + y = 2, 0.4 } \}$.

The \emph{preference function} induced by an SCSP $P = \angl{V, E, C}$, with $C = \{ \angl{U_1, E, {[\cdot]}_1}, \dots, \angl{U_n, E, {[\cdot]}_n} \}$ is the function ${[\cdot]}_P: \Assign(V) \to E$ defined by ${[\alpha]}_P = {[{\alpha\restriction_{U_1}}]}_1 \otimes \dots \otimes {[\alpha\restriction_{U_n}]}_n$ when $n > 0$ and ${[\alpha]}_P = \boldone$ otherwise. It is not hard to see that every SCSP can be reduced to have exactly one SDC, namely $\angl{V, E, {[\cdot]}_P}$. The advantage of SCSPs containing multiple SDCs is that they can often express preferences more concisely by only considering values of symbols relevant to the concern at hand.

\begin{definition}[SCSP solutions]
\label{def:scsp-solution}
Let $P = \angl{V, E, C}$ be an SCSP\@. A \emph{solution} to $P$ is an assignment $\alpha$ of $V$ such that ${[\alpha]}_P \neq \boldzero_E$ and there exists no other assignment $\beta$ of $V$ with ${[\alpha]}_P <_E {[\beta]}_P$. We write $\Sol(P)$ for the set of solutions to $P$.
\end{definition}

For the example SCSP $K$ mentioned above, $\alpha \in \Assign(\{ x, y\})$ such that $\alpha(x) = 0.5$ and $\alpha(y) = 1.5$, with preference value $0.9 \cdot{} 0.4 = 0.36$ qualifies for a solution.

Observe that the wording of Definition~\ref{def:scsp-solution} accommodates for the possibility that $\leq_E$ may not be total. We can also compose SCSPs that share an underlying semiring, simply by taking the union of their variables and SDCs.\footnote{Similar to~\cite{bistarelli-2004}, we assume that no two SCSPs being composed share the exact same SDC.}

\begin{definition}[SCSP composition]
Let $P_1 = \angl{V_1, E, C_1}$ and $P_2 = \angl{V_2, E, C_2}$ be SCSPs. Their \emph{composition}, written $P_1 \otimes P_2$, is the SCSP $\angl{V_1 \cup V_2, E, C_1 \cup C_2}$.
\end{definition}

By this definition, we immediately see that the operator $\otimes$ on SCSPs is commutative and associative. At this point, it is worth noting that classical CSPs are captured by SCSPs; by the definitions above, a CSP is simply an SCSP over the boolean semiring. Conversely, one can translate an SCSP into a CSP by simply having a single constraint encode the requirement for a solution given in Definition~\ref{def:scsp-solution}.

Lastly, we can move SCSPs between c-semirings using homomorphisms:
\begin{definition}[homomorphisms for SCSPs]
Let $P = \angl{V, E_1, C}$ be an SCSP and $h: E_1 \to E_2$ be a homomorphism. Then $h(P)$ is the SCSP $\angl{V, E_2, C'}$ where $C' = \{ \angl{U, E_2, h \circ {[\cdot]}} : \angl{U, E_1, [\cdot]} \in C \}$.
\end{definition}

One can prove~\cite{li-ying-2008} that if $h$ is order-reflecting, then $h$ will preserve the solutions to an SCSP, or more precisely, $\Sol(P) = \Sol(h(P))$. Moreover, it follows immediately from the above definition that homomorphisms are compatible with SCSP composition, in the sense that if $P_1$ and $P_2$ are SCSPs that have $E$ as underlying semiring and $h: E \to E'$ is a homomorphism, then $h(P_1 \otimes P_2) = h(P_1) \otimes h(P_2)$.

\subsection{Separating SCSPs from CSPs}
\label{sec:scsps-vs-csps}

A watershed between SCSPs and CSPs becomes clear when we consider the composition operator. Let $P_1 = \angl{V_1, \mathbb{B}, C_1}$ and $P_2 = \angl{V_2, \mathbb{B}, C_2}$ be CSPs. If $\alpha$ is a solution to $P_1 \otimes P_2$, then $\alpha\restriction_{V_i}$ is a solution to $P_i$ for $i \in \{1,2\}$. Because of the way preferences compose, this property need not hold when the semiring is something other than the boolean semiring. Loss of this property can be useful: it exhibits the possibility of SCSPs \emph{compromising} when higher-preference assignments turn out to be incompatible.\footnote{Conversely, if $\alpha_1$ and $\alpha_2$ are solutions to CSPs $P_1$ and $P_2$ respectively, and both agree on common symbols, then $\alpha_1 + \alpha_2$ is necessarily a solution to $P_1 \otimes P_2$. This property also does not hold for SCSPs; we refer to the technical report~\cite{techreport} for details.}


\section{Components}
\label{sec:components}


We now introduce Soft Constraint Automata, as the most fundamental building blocks of our framework. Soft Constraint Automata were first proposed as a generalization of Constraint Automata~\cite{baier-sirjani-arbab-rutten-2006} in~\cite{arbab-santini-2013}, to enable service discovery based on non-crisp preferences. In contrast, Soft Constraint Automata as used in this paper employ their preferences purely as a means to select their next transition.

\begin{definition}[Soft Constraint Automaton]
A \emph{Soft Constraint Automaton} (SCA) is a tuple $\angl{Q, V, E, \rightarrow, q^0}$ where
\begin{itemize}
    \setlength{\itemsep}{0em}
    \item $Q$ is a finite set of states, with $q^0 \in Q$
    \item $V$ is a finite set of port symbols
    \item $E$ is a c-semiring, referred to as the \emph{underlying semiring}
    \item $\rightarrow\ \subseteq Q \times \SCSP(V, E) \times Q$, the \emph{transition relation}
\end{itemize}
\end{definition}

When $\angl{q, P, q'} \in\ \rightarrow$, we write $q \xrightarrow{P} q'$; if $P$ contains only the binary constraint $\angl{U, \phi, e}$, we write $q \xrightarrow{U,\ \phi,\ e} q'$. Note that, unlike~\cite{arbab-santini-2013}, we truly label transitions with SCSPs rather than a set of ports and a single data constraint (the set of ports is implied by the SCSP labels in our notation). Analogous to SCSPs, one can observe that because Constraint Automata (CAs)~\cite{baier-sirjani-arbab-rutten-2006} have their transitions labeled with CSPs (which are simply SCSPs over the boolean c-semiring), they can be obtained as SCAs with the boolean semiring as underlying semiring~\cite{arbab-santini-2013}.

We can also lift homomorphisms to operate on SCAs, as they did on SCSPs.
\begin{definition}[homomorphisms for SCAs]
\label{def:sca-homomorphism}
Let $A = \angl{Q, V, E, \rightarrow, q^0}$ be an SCA and let $h: E \to E'$ be a homomorphism. Then $h(A)$ is the SCA $\angl{Q, V, E', \rightarrow_h, q^0}$, where $\rightarrow_h$ is the smallest relation such that $q \xrightarrow{h(P)}_h q'$ whenever $q \xrightarrow{P} q'$.
\end{definition}

The semantics of an SCA is obtained by means of its execution model, as the stream of solutions to the SCSPs that label the transitions. As a consequence, each element of this stream is a (partial) assignment of the set of port symbols of the automaton, $V$.
\begin{definition}[SCA semantics]
Let $A = \angl{Q, V, E, \rightarrow, q^0}$ be an SCA\@. Its \emph{execution relation} is the smallest binary relation $\Rightarrow_A$ on $Q^\omega \times {\Assign_\subseteq(V)}^\omega$ satisyfing the rule
$$
\inferrule{%
    \lambda(0) \xrightarrow{P} \lambda(1) \\
    \pi(0) \in \Sol(P)
}{%
    \angl{\lambda, \pi} \Rightarrow_A \angl{\tail(\lambda), \tail(\pi)}
}
$$
The \emph{language} accepted by $A$, written $L(A)$, consists of the elements $\pi \in {\Assign_\subseteq(V)}^\omega$ such that there exists a stream $\lambda \in Q^\omega$ with $\lambda(0) = q^0$ and $\angl{\lambda, \pi} \Rightarrow_A \angl{\tail(\lambda), \tail(\pi)} \Rightarrow_A \angl{\tail^2(\lambda), \tail^2(\pi)} \Rightarrow_A \dots$
\end{definition}

It is easy to observe that for any SCA, $A$, there exists an SCA $A'$ with the boolean semiring as its underlying semiring, such that $\Rightarrow_A$ is the same relation as $\Rightarrow_{A'}$, and, by extension, $L(A) = L(A')$ (refer to Section~\ref{sec:example-movement} for an example). Moreover, if $h$ is an order-reflecting homomorphism, one can quickly see that $L(A) = L(h(A))$. By the latter observation, we can consider a CA to be an SCA over any c-semiring $E$, as finding an order-reflecting homomorphism $h: \mathbb{B} \to E$ is trivial: simply map the units of $\mathbb{B}$ to those of $E$. In the sequel, we therefore regard a CA as an SCA over some semiring $E$ whenever convenient.

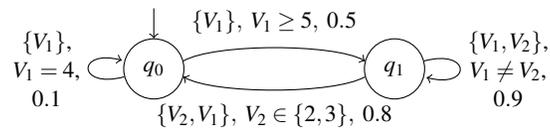
\begin{figure}[h!]
\centering
\begin{tikzpicture}[scale=\scale, transform shape]
    \begin{scope}[
        every node/.style={draw,circle},
        minimum size=\nodesize
    ]
        \node (0) at (0,0) {$q_0$};
        \node (1) at (4,0) {$q_1$};
    \end{scope}

    \node[above=5mm of 0] (s) {};
    \path[->] (s) edge (0);

    \begin{scope}[
        every path/.style={->,looseness=0.5},%
        every node/.style={align=center}%
    ]
        \path (0.\angle) edge[bend left] node[above=1mm] {$\{V_1\},\ V_1 \geq 5,\ 0.5$} (1.180-\angle);
        \path (1.180+\angle) edge[bend left] node[below=1mm] {$\{V_2,V_1\},\ V_2 \in \{2,3\},\ 0.8$} (0.-\angle);
        \path (0) edge[loop left] node[left] {$\{V_1\}$,\\ $V_1 = 4$,\\ $0.1$} (0);
        \path (1) edge[loop right] node[right] {$\{V_1,V_2\}$,\\ $V_1 \neq V_2$,\\ $0.9$} (1);
    \end{scope}
\end{tikzpicture}
\caption{An example of our graphical representation of SCAs}\label{fig:example-sca}
\end{figure}

Lastly, a graphical representation of SCAs is appropriate for the remainder of this paper. We depict SCAs as transition systems; the labels on the edges of our transition systems will in all cases be the abbreviated representation of binary SDCs mentioned in the previous section. Because we restrict graphical depictions to use this abbreviation, we will not be able to draw all SCAs, but the ones we can draw will have less convoluted representations and will suffice for our purposes in Section~\ref{sec:example}. For an example of an SCA with the probabilistic c-semiring $\mathbb{P}$ as the underlying semiring, refer to Figure~\ref{fig:example-sca}. We note that, in this representation, the set of port symbols of the automaton, $V$, is left implicit as the union of the occuring variable sets; i.e., in Figure~\ref{fig:example-sca}, $V = \{V_1, V_2\}$.

\section{Composition}
\label{sec:composition}


We now turn our attention to composition of SCAs. Firstly and most obviously, we can lift the composition operator of CAs~\cite{baier-sirjani-arbab-rutten-2006} to SCAs, as is done in~\cite{arbab-santini-2013}. Intuitively, this composition operator composes transitions that agree on common ports. A side-condition here is that the operands have the same underlying semiring.

\begin{definition}[product composition of SCAs]
\label{def:sca-product-composition}
Let $A_1 = \angl{Q_1, V_1, E, \rightarrow_1, q^0_1}$ and $A_2 = \angl{Q_2, V_2, E, \rightarrow_2, q^0_2}$ be two SCAs. Their \emph{product composition}, written $A_1 \otimes A_2$, is the SCA $\angl{Q_1 \times Q_2, V_1 \cup V_2, E, \rightarrow, \angl{q^0_1, q^0_2}}$, in which $\rightarrow$ is the smallest relation satisyfing the rule
$$
\inferrule{%
    q_1 \xrightarrow{\angl{V_1',\ E,\ C_1}}_1 q_1' \\
    q_2 \xrightarrow{\angl{V_2',\ E,\ C_2}}_2 q_2' \\
    V_1' \cap V_2 = V_2' \cap V_1
}{%
    \angl{q_1, q_2} \xrightarrow{\angl{V_1',\ E,\ C_1}\ \otimes\ \angl{V_2',\ E,\ C_2}} \angl{q_1', q_2'}
}
$$
\end{definition}

By the above definition, it is obvious that the $\otimes$-operator for SCAs is commutative and associative, modulo a simple relabeling of the states. In the sequel, we abstract from this relabeling, as it has no bearing on the semantics of the automaton. Furthermore, application of a homomorphism to SCAs (as in Definition~\ref{def:sca-homomorphism}) is compatible with product composition in the sense that for SCAs $A_1$ and $A_2$ that share an underlying semiring $E$ and a homomorphism $h: E \to E'$, we have $h(A_1 \otimes A_2) = h(A_1) \otimes h(A_2)$.

One can now see that the same observations that distinguish composition of SCSPs from that of CSPs in Section~\ref{sec:scsps-vs-csps}, also distinguish composition of SCAs from that of CAs. At this level, the advantage of these properties for describing Cyber-Physical Systems is more clear; when the transitions preferred by the operands $A_1$ and $A_2$ are incompatible, $A_1 \otimes A_2$ may still have transitions composed of lower-preference transitions found in $A_1$ and $A_2$. Informally, one may say that by providing alternative (lower-preference) transitions, SCAs can be made resillient in their composition with unforeseen automata.

One notable difference of Definition~\ref{def:sca-product-composition} with the definition of composition for CAs in~\cite{arbab-santini-2013} is that we exclude rules for independent transitions, i.e., those of the form
\begin{equation}
\label{rule:indep}
\inferrule{%
    q_i \xrightarrow{\angl{V_i',\ E,\ C_i}}_i q_i' \\
    q_j \in Q_j \\
    V_i' \cap V_j = \emptyset
}{%
    \angl{q_i, q_j} \xrightarrow{\angl{V_i',\ E,\ C_j}} \angl{q_i', q_j}
}
\end{equation}
This is due to the fact that, if these rules are included, the product composition will be biased towards independent transitions in terms of preference (due to the fact that for $e_1, e_2 \in E$ we have $e_1 \otimes e_2 \leq_E e_1$; c.f.~\cite[Theorem~2.1.3]{bistarelli-2004}), unless the preference value of the original transition is composed with some other preference value $e$. If we choose $e = \boldone_E$ for this preference value, the composed preference is unchanged (and we are essentially applying~\eqref{rule:indep} as-is). If on the other hand we choose $e = \boldzero_E$, then we are in fact prohibiting independent transitions, as the composed preference value will be $\boldzero_E$. Having established that neither $\boldzero$ nor $\boldone$ is a suitable candidate, we leave it to the designer of the SCA to include self-transitions, each with an appropriate preference value $e \in E$, to states of an automaton where other automata are permitted to make independent transitions, and vice versa. For a concrete use of this technique, we refer to Section~\ref{sec:example}. Note that $e$ need not be the same for all states; a component may have a different preference regarding independent moves by other components depending on context. For example, it is conceivable that in some situations, a component may want to (almost) completely inhibit behavior of other components until some task is completed.

The product composition operator of Definition~\ref{def:sca-product-composition} to CAs now coincides with the product composition operator for CAs of~\cite{baier-sirjani-arbab-rutten-2006}, provided every state $q$ has a self-transition $q \xrightarrow{\emptyset,\ \top,\ \boldone} q'$~\cite{jongmans-kappe-arbab-2015}.

Of course, it may also occur that we want to compose SCAs that model different concerns. In this case, it may not make sense to use the product composition operator proposed above. Moreover, if the preferences of the operands are expressed in different semirings, the product composition operator in does not apply. To deal with such cases, we introduce another variant of composition, based on the join product. However, for such an operator to make sense, we need to be able to encode the preferences expressed by the operands into the composed preference domain. Therefore, we must add a side condition guaranteeing that such an embedding is possible.

\begin{definition}[join composition of SCAs]
Let $A_1$ and $A_2$ be SCAs with cancellative underlying semirings $E_1$ and $E_2$ respectively.
Their \emph{join composition}, written $A_1 \odot A_2$, is the SCA $h_L(A_1) \otimes h_R(A_2)$, in which $h_L$ and $h_R$ are the canonical injections from $E_1$ to $E_1 \odot E_2$ and $E_2$ to $E_1 \odot E_2$ respectively.
\end{definition}

The operator $\odot$ on SCAs is also commutative and associative, up to a trivial order-reflecting homomorphism, i.e., if $A_1$, $A_2$ and $A_3$ are SCAs, then $A_1 \odot A_2 = h(A_2 \odot A_1)$ and $A_1 \odot (A_2 \odot A_3) = h'((A_1 \odot A_2) \odot A_3)$ for some order-reflecting homomorphisms $h$ and $h'$. In the sequel, we abstract from this homomorphism and simply regard $\odot$ as commutative and associative. Going by the fact that $A_1 \odot A_2$ has $E_1 \odot E_2$ as underlying semiring, we can surmise that the transitions allowed in $A_1 \odot A_2$ are those that are Pareto-optimal with respect to the preferences of both automata. In other words, the join composition of two SCAs will exclude behavior that ``unnecessarily inconveniences'' one of the operands.

We can also use the lexicographic composition described in Definition~\ref{def:lexicographic-product-semiring} to compose automata where the preferences expressed by one are subsidiary to the other. Again, we must include a side-condition guaranteeing that we can embed the preferences from the operands.
\begin{definition}[lexicographic composition of SCAs]
Let $A_1$ and $A_2$ be SCAs with underlying cancellative c-semirings $E_1$ and $E_2$ respectively. Their \emph{lexicographic composition}, written $A_1 \triangleright A_2$, is the SCA $h_L(A_1) \otimes h_R(A_2)$, in which $h_L$ and $h_R$ are the canonical injections from $E_1$ and $E_2$ to $E_1 \triangleright E_2$ respectively.
\end{definition}

We observe that, unlike $\otimes$ and $\odot$, the operator $\triangleright$ is (by design) not commutative. However, it is associative, in part due to the fact that if c-semirings $E_1$ and $E_2$ are cancellative, then so is $E_1 \triangleright E_2$.

\section{Example: Patrolling Agent}
\label{sec:example}


In this example, we consider an agent that is tasked with patrolling along a predefined path. We start with the simplest description possible and subsequently extend our model to include deviation from the path as well as energy usage. At every step, we describe the environment in terms of Constraint Automata, after which we express the relevant preferences of the agent using Soft Constraint Automata.

Composite states are denoted by (flattened) tuples; i.e., if $p$, $q$ and $r$ are states of the automata $A_1$, $A_2$ and $A_3$ respectively, then $\angl{p, q, r}$ is a state of the automaton $A_1 \odot A_2 \odot A_3$. As a special case, we omit the state symbol of single-state automata from the state tuple; their sole state is also written as $\epsilon$ to reflect this. Finally, all c-semirings used in this section are tacitly assumed to be cancellative.

\subsection{Movement back and forth}
\label{sec:example-movement}

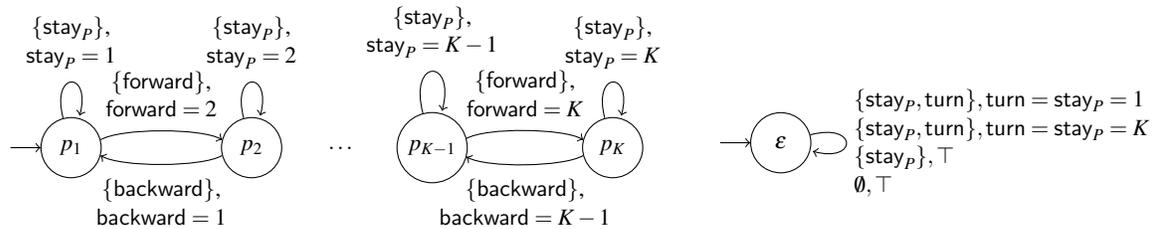
\begin{figure}[h]
\begin{subfigure}[b]{0.6\textwidth}
\centering
\begin{tikzpicture}[scale=\scale, transform shape]
    \begin{scope}[every node/.style={draw,circle},minimum size=\nodesize]
        \foreach \x/\n in {0/1, 1/2, 2/K-1, 3/K}{%
            \node (\n) at (\x*\gap,0) {$p_{\n}$};
        }
    \end{scope}
    \node at (4.5, 0) {$\dots$};

    \node[left=5mm of 1] (s) {};
    \path[->] (s) edge (1);

    \begin{scope}[%
        every path/.style={->,looseness=0.5},
        every node/.style={align=center}
    ]
        \foreach \src/\dst in {1/2, K-1/K}{%
            \path (\src.\angle) edge[bend left] node[above=1mm] {$\{\mforward\}$,\\ $\mforward = \dst$} (\dst.180-\angle);
            \path (\dst.180+\angle) edge[bend left] node[below=1mm] {$\{\mbackward\}$,\\ $\mbackward = \src$} (\src.-\angle);
        }

        \foreach \n/\pos in {1/above, 2/above, K-1/above, K/above}{%
            \path (\n) edge[loop \pos] node[\pos=1mm] {$\{\mstay_P\}$,\\ $\mstay_P = \n$} (\n);
        }
    \end{scope}
\end{tikzpicture}
\caption{A sketch of the CA $A_\mathsf{path}$, modelling the position along the path.}\label{fig:ca-path}
\end{subfigure}
\begin{subfigure}[b]{0.39\textwidth}
\centering
\begin{tikzpicture}[scale=\scale, transform shape]
    \begin{scope}[every node/.style={draw,circle},minimum size=\nodesize]
        \node (e) at (0,0) {$\epsilon$};
    \end{scope}

    \node[left=5mm of e] (s) {};
    \path[->] (s) edge (e);

    \begin{scope}[every path/.style={->}, every node/.style={align=left}]
        \path (e) edge[loop right] node[right] {%
            $\{ \mstay_P, \turn \}, \turn = \mstay_P = 1$\\
            $\{ \mstay_P, \turn \}, \turn = \mstay_P = K$\\
            $\{ \mstay_P \}, \top$ \\
            $\emptyset, \top$
        } (e);
    \end{scope}
\end{tikzpicture}
\vspace{4mm}
\caption{The CA $A_{\turn,P}$, modelling turning points.}\label{fig:ca-turn-p}
\end{subfigure}
\caption{CAs pertaining to the path.}
\end{figure}

The path of the agent is broken up into $K$ discrete positions, each of which is represented by a state; positions $1$ and $K$ are the endpoints of the path. At every position, the agent may move forward or backward along the path. The agent can also opt to stay at its current position; these self-transitions allow other components to make transitions independently, i.e., make moves while the agent remains stationary. The constraint automaton $A_\mathsf{path}$, which represents these possibilities, is given in Figure~\ref{fig:ca-path}. To provide a turning point for the agent, we use the CA $A_{\turn,P}$ in Figure~\ref{fig:ca-turn-p}; note that in $A_\mathsf{patrol} \otimes A_{\turn,P}$ the agent is required to stay put when turning: if $\mstay_P$ does not fire, then neither can $\turn$.

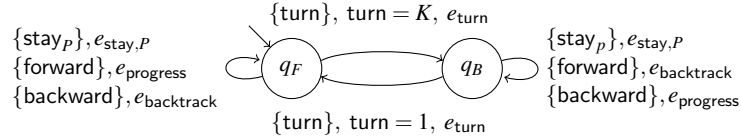
\begin{figure}[h]
\centering
\begin{tikzpicture}[scale=\scale, transform shape]
    \begin{scope}[every node/.style={draw,circle},minimum size=\nodesize]
        \node (F) at (0,0) {$q_F$};
        \node (B) at (3,0) {$q_B$};
    \end{scope}

    \node[above left=5mm of F] (s) {};
    \path[->] (s) edge (F);

    \begin{scope}[%
        every path/.style={->,looseness=0.5},%
        every node/.style={align=left}%
    ]
        \path (F.\angle) edge[bend left] node[above=3mm] {%
            $\{\turn\},\ \turn = K,\ e_\turn$
        } (B.180-\angle);
        \path (B.180+\angle) edge[bend left] node[below=3mm] {%
            $\{\turn\},\ \turn = 1,\ e_\turn$
        } (F.-\angle);
        \path (F) edge[loop left] node[left] {%
            $\{\mstay_P\}, e_{\mstay,P}$ \\
            $\{\mforward\}, e_\progress$ \\
            $\{\mbackward\}, e_\backtrack$
        } (F);
        \path (B) edge[loop right] node[right] {%
            $\{\mstay_p\}, e_{\mstay,P}$ \\
            $\{\mforward\}, e_\backtrack$ \\
            $\{\mbackward\}, e_\progress$
        } (B);
    \end{scope}
\end{tikzpicture}
\caption{The SCA $A_\mathsf{patrol}$, expressing patrolling preferences.}\label{fig:sca-patrol}
\end{figure}

The preference of the agent to keep patrolling is expressed by the two-state SCA in Figure~\ref{fig:sca-patrol}. In this description, the transitions are labeled with preferences from some semiring $E$. To ensure that the agent does indeed patrol, we stipulate that $e_\backtrack <_E e_{\mstay,P} <_E e_\turn, e_\progress$. In other words, the agent will prefer to start another lap or progress (in its current target direction); failing that, it would rather stay at its current position than backtrack to the previous one.

The final system as presented up to this point is now formed by the following expression (recall that we can interpret the CAs $A_\mathsf{path}$ and $A_{\turn, P}$ as SCAs with the same underlying semiring as $A_\mathsf{patrol}$):
$$A_\mathsf{move} = A_\mathsf{path} \otimes A_{\turn,P} \otimes A_\mathsf{patrol}$$
By itself, this automaton is not particularly interesting; it will exhibit precisely the desired patrolling behavior. As a matter of fact, it is almost trivial to use a Constraint Automaton and achieve the exact same behavior, by pruning the transitions whose preference is dominated by that of some other transition, and setting the preference value of the remaining transitions to $\boldone$. In this case, this would mean that we delete:
\begin{itemize}
    \setlength{\itemsep}{0em}
    \item the transitions involving $e_{\mstay,P}$ and $e_\backtrack$ in $A_\mathsf{patrol}$ from both states
    \item the self-transition that fires $\mbackward$ (respectively $\mforward$) from $q_F$ (respectively $q_B$) from $A_\mathsf{patrol}$
    \item the self-transition with $\mstay_P$ in states $p_2,\dots,p_{K-1}$ from $A_\mathsf{path}$.
\end{itemize}
We will see in the sequel that allowing these ``alternative'' transitions to remain enables us to extend our model more easily later on.
\subsection{Deviating from the path}

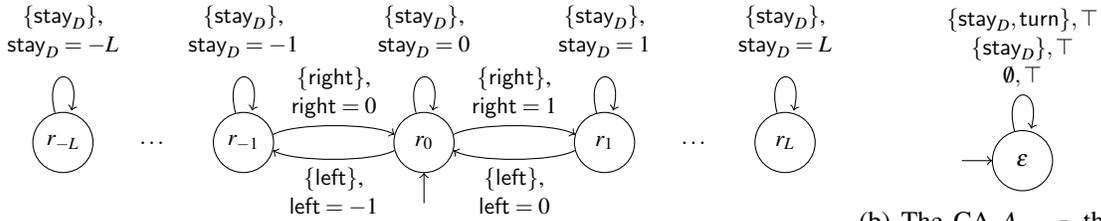
\begin{figure}[h!]
\begin{subfigure}[b]{0.72\textwidth}
\centering
\begin{tikzpicture}[scale=\scale, transform shape]
    \begin{scope}[every node/.style={draw,circle},minimum size=\nodesize]
        \foreach \x/\n in {-2/-L, -1/-1, 0/0, 1/1, 2/L}{%
            \node (\n) at (\x*\gap,0) {$r_{\n}$};
        }
    \end{scope}
    \foreach \after in {-2, 1}{%
        \node at ({(\after+1/2)*\gap}, 0) {$\dots$};
    }

    \node[below=5mm of 0] (s) {};
    \path[->] (s) edge (0);

    \begin{scope}[%
        every path/.style={->,looseness=0.5},
        every node/.style={align=center},
    ]
        \foreach \src/\dst in {0/1, -1/0}{%
            \path (\src.\angle) edge[bend left] node[above=0mm] {$\{\mright\}$,\\ $\mright = \dst$} (\dst.180-\angle);
            \path (\dst.180+\angle) edge[bend left] node[below=0mm] {$\{\mleft\}$,\\ $\mleft = \src$} (\src.-\angle);
        }

        \foreach \n in {-L,-1,0,1,L}{%
            \path (\n) edge[loop above] node[above=2mm] {$\{\mstay_D\}$,\\ $\mstay_D = \n$} (\n);
        }
    \end{scope}
\end{tikzpicture}
\caption{A sketch of the CA $A_\mathsf{stray}$, modelling deviation from the path in any position.}\label{fig:ca-deviation}
\end{subfigure}
\begin{subfigure}[b]{0.27\textwidth}
\centering
\begin{tikzpicture}[scale=\scale, transform shape]
    \begin{scope}[every node/.style={draw,circle},minimum size=\nodesize]
        \node (e) at (0,0) {$\epsilon$};
    \end{scope}

    \node[left=5mm of e] (s) {};
    \path[->] (s) edge (e);

    \begin{scope}[every path/.style={->}, every node/.style={align=center}]
        \path (e) edge[loop above] node[above] {%
            $\{ \mstay_D, \turn \}, \top$ \\
            $\{ \mstay_D \}, \top$ \\
            $\emptyset, \top$
        } (e);
    \end{scope}
\end{tikzpicture}
\caption{The CA $A_{\turn,D}$ that fixes the turning point on the path.}\label{fig:ca-turn-d}
\end{subfigure}
\caption{CAs pertaining to deviation from the path.}
\end{figure}

We now add in the possibility for our agent to stray from its predetermined path to the left or the right by $L$ steps, while still preferring that the agent stays on the path. To model this in the environment, we construct the CA sketched in Figure~\ref{fig:ca-deviation}. In this CA, the state $r_0$ represents that the agent is on the path, while the other states $r_i$ represent that the agent is located $|i|$ steps beside the current position on the path towards the next turning point, with the sign of $i$ indicating the direction of divergence (negative for deviation to the left, positive otherwise).

We now encourage the reader to consider what the automaton $A_\mathsf{path} \otimes A_\mathsf{deviate}$ (c.f. Figures~\ref{fig:ca-path} and~\ref{fig:ca-deviation}) looks like, as it will be a subcomponent of our final system. For instance, when this automaton is in state $\angl{p_i, q_j}$, with $i < n$ and $j < L$, it can transition to state $\angl{p_{i+1}, q_{j+1}}$ by firing $\{\mforward, \mright\}$.

Moving on, we need the CA in Figure~\ref{fig:ca-turn-d} to make sure that the agent cannot turn and diverge (or turn and converge) at the same time. If we want to require that the agent turns only when it is on the path, we can add the constraint $\mstay_D = 0$ to the self-transition that fires $\turn$ in $A_{\turn,D}$.

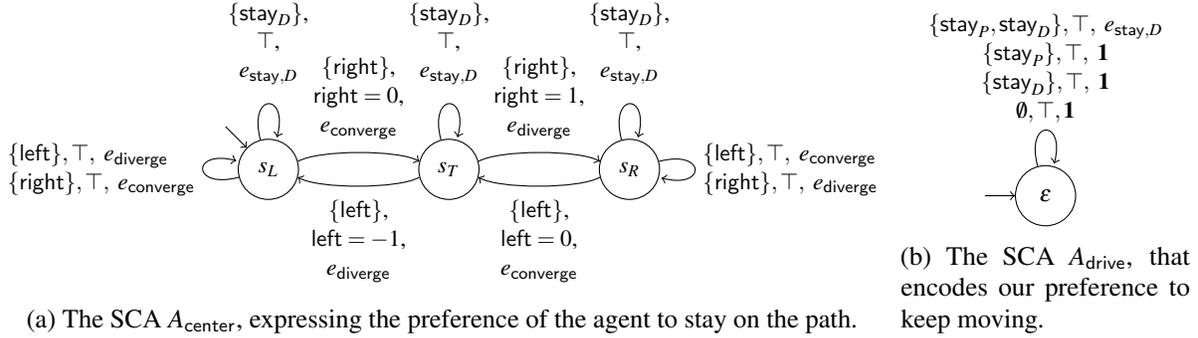
\begin{figure}[h!]
\begin{subfigure}[b]{0.75\textwidth}
\centering
\begin{tikzpicture}[scale=\scale, transform shape]
    \begin{scope}[every node/.style={draw,circle},minimum size=\nodesize]
        \foreach \x/\n in {-1/L, 0/T, 1/R}{%
            \node (\n) at (\x*\gap,0) {$s_{\n}$};
        }
    \end{scope}

    \node[above left=5mm of L] (s) {};
    \path[->] (s) edge (L);

    \begin{scope}[%
        every path/.style={->,looseness=0.5},%
        every node/.style={align=center}%
    ]
        \foreach \src/\dst/\labelone/\labeltwo in {0/1/diverge/converge, -1/0/converge/diverge}{%
            \path (\src.\angle) edge[bend left] node[above=1mm] {$\{\mright\}$,\\ $\mright = \dst$,\\ $e_\mathsf{\labelone}$} (\dst.180-\angle);
            \path (\dst.180+\angle) edge[bend left] node[below=1mm] {$\{\mleft\}$,\\ $\mleft = \src$,\\ $e_\mathsf{\labeltwo}$} (\src.-\angle);
        }
        \begin{scope}[%
            every node/.style={align=left}%
        ]
            \path (L) edge[loop left] node[left] {%
                $\{\mleft\},\top,\ e_\diverge$ \\
                $\{\mright\},\top,\ e_\converge$
            } (L);
            \path (R) edge[loop right] node[right] {%
                $\{\mleft\},\top,\ e_\converge$ \\
                $\{\mright\},\top,\ e_\diverge$
            } (R);
        \end{scope}

        \foreach \n in {L,T,R}{%
            \path (\n) edge[loop above] node[above=2mm] {$\{\mstay_D\}$,\\ $\top$,\\ $e_{\mstay,D}$} (\n);
        }
    \end{scope}
\end{tikzpicture}
\caption{The SCA $A_\mathsf{center}$, expressing the preference of the agent to stay on the path.}\label{fig:sca-center}
\end{subfigure}
\begin{subfigure}[b]{0.24\textwidth}
\centering
\begin{tikzpicture}[scale=\scale, transform shape]
    \begin{scope}[every node/.style={draw,circle,minimum size=\nodesize}]
        \node (e) {$\epsilon$};
    \end{scope}

    \node[left=5mm of e] (s) {};
    \path[->] (s) edge (e);

    \begin{scope}[%
        every path/.style={->,looseness=0.5},%
        every node/.style={align=center}%
    ]
        \path (e) edge[loop above] node[above] {%
            $\{ \mstay_P, \mstay_D \},\top,\ e_{\mstay,D}$ \\
            $\{ \mstay_P \},\top,\ \boldone$ \\
            $\{ \mstay_D \},\top,\ \boldone$ \\
            $\emptyset, \top, \boldone$
        } (e);
    \end{scope}
\end{tikzpicture}
\caption{The SCA $A_\mathsf{drive}$, that encodes our preference to keep moving.}\label{fig:sca-drive}
\end{subfigure}
\caption{SCAs pertaining to movement preferences.}
\end{figure}

As indicated, we prefer for the agent to stay on the path as much as possible. The preferences expressed to this end are of a different form than those of Figure~\ref{fig:sca-patrol}; after all, when the agent has strayed from the path to the right, we prefer to go to the left, and vice versa. If the agent is still on the path, it prefers to remain that way. Such preferences are expressed in the SCA in Figure~\ref{fig:sca-center}, where we require that $e_\diverge <_E e_{\mstay,D} <_E e_\converge$. In this SCA, state $s_L$ represents that the agent has deviated to the left, while $s_R$ models a deviation to the right. We can see that in $s_T$ (where the agent is on the path), the most preferred transition is the self-transition firing $e_{\mstay,D}$.

To encode the preference that the agent should not stay in the exact same state in both $A_\mathsf{stray}$ and $A_\mathsf{path}$ (i.e., it is better to diverge than not to move at all), we compose $A_\mathsf{center}$ with the SCA depicted in Figure~\ref{fig:sca-drive}. This SCA expresses its preferences over the same semiring $E$ as $A_\mathsf{center}$, and we postulate that $e_{\mstay,D} \otimes e_{\mstay,D} <_E e_\diverge$. Informally, this SCA penalizes firing port $\mstay_D$ in concert with port $\mstay_P$ by the additional ``cost'' $e_{\mstay,D}$ and leaves all other preferences intact.

By expressing this preference in a separate SCA we achieve a (desirable) separation of the concern ``the agent should stay on the path'' from ``the agent should keep moving''. More importantly, however, by creating a separate SCA for the latter concern we also save ourselves the error-prone effort of manually working out the possible combinations of firing $\mright$ with or without firing $\mstay_D$ concurrently.

The component that describes deviation from the path is now given by
$$A_\mathsf{deviate} = A_\mathsf{stray} \otimes A_{\turn,D} \otimes A_\mathsf{center} \otimes A_\mathsf{drive}$$
Now, to compose $A_\mathsf{move}$ with $A_\mathsf{deviate}$, we have several options. First of all, if they share their underlying semiring, we can simply calculate their product composition $A_\mathsf{move} \otimes A_\mathsf{deviate}$. There is, however, an objection to using the product composition operator in this particular case. Unless we tune our preferences carefully, we may find that $e_\progress \otimes e_{\mstay,D} <_E e_{\mstay,P} \otimes e_\converge$. If our primary concern is for the agent to patrol, then such a preference may be undesirable, for it will cause the agent to choose returning to the path over continuing its patrol beside the path.

Since patrolling and staying on the path are separate concerns, one may propose to use the join composition on these components, i.e., $A_\mathsf{move} \odot A_\mathsf{deviate}$.  However, this causes $\angl{e_\progress, e_{\mstay,D}}$ and $\angl{e_{\mstay,P}, e_{\converge}}$ be unordered, meaning that in a state where both transitions are available, neither would be preferred over the other. Indeed, the only method to enforce the importance of one concern over the other is to use the lexicographic join operator and calculate $A_\mathsf{move} \triangleright A_\mathsf{deviate}$.

To illustrate the resulting automaton, we consider the following transitions:\setcounter{equation}{0}
\begin{align}
\label{transition:progress-converge}
\angl{p_1, q_F, r_{-1}, s_L} \xrightarrow{\{\mforward,\ \mright\},\, \mright = 0\ \wedge\ \mforward = 2} \angl{p_2, q_F, r_0, s_T} \\
\label{transition:progress-stay}
\angl{p_1, q_F, r_{-1}, s_L} \xrightarrow{\{\mforward,\ \mstay_D\},\, \mstay_D = -1\ \wedge\ \mforward = 2} \angl{p_2, q_F, r_{-1}, s_L} \\
\label{transition:stay-diverge}
\angl{p_n, q_B, r_0, s_T} \xrightarrow{\{\mstay_P,\ \mright\},\, \mstay_P = L \ \wedge\ \mleft = 1} \angl{p_n, q_B, r_1, s_R} \\
\label{transition:stay-stay}
\angl{p_n, q_B, r_0, s_T} \xrightarrow{\{\mstay_P,\ \mstay_D \},\, \mstay_P = L \ \wedge\ \mstay_D = 0} \angl{p_n, q_B, r_0, s_T}
\end{align}
We can see that transition~\eqref{transition:progress-stay} is preferred over transition~\eqref{transition:progress-converge} since $\angl{e_\progress, e_{\mstay,D}}$ is a value preferred over $\angl{e_\progress, e_\converge}$. This means that our agent would rather move forward and converge at the same time than just move forward. Also, transition~\eqref{transition:stay-diverge} is preferred over transition~\eqref{transition:stay-stay} for $\angl{e_{\mstay,P}, e_{\mstay,D} \otimes e_{\mstay,D}} <_E \angl{e_{\mstay,P}, e_\diverge}$; we can take this to mean that, if the agent cannot progress, it prefers to deviate from its path (and possibly find a path forward) over staying in the same location.

As a final observation in this section, we note that because $A_\mathsf{drive}$ is in the second component of the lexicographical composition, the agent is still allowed to turn while stationary at the endpoints of the path.

\subsection{Energy usage}

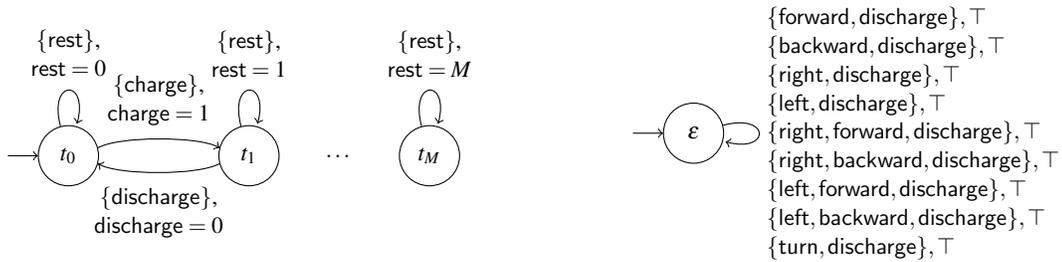
\begin{figure}[h]
\begin{subfigure}[b]{0.55\textwidth}
\centering
\begin{tikzpicture}[scale=\scale, transform shape]
    \begin{scope}[%
        every node/.style={draw,circle,minimum size=\nodesize}%
    ]
        \foreach \x/\n in {0/0, 1/1, 2/M}{%
            \node (\n) at (\x*\gap, 0) {$t_{\n}$};
        }
    \end{scope}

    \node[left=5mm of 0] (s) {};
    \path[->] (s) edge (0);

    \begin{scope}[%
        every node/.style={align=center}%
    ]
        \begin{scope}[%
            every path/.style={->,looseness=0.5,bend left}%
        ]
            \foreach \src/\dst in {0/1}{%
                \path (\src.\angle) edge node[above=1mm] {$\{ \charge \}$,\\ $\charge = \dst$} (\dst.180-\angle);
                \path (\dst.180+\angle) edge node[below=1mm] {$\{ \discharge \}$,\\ $\discharge = \src$} (\src.-\angle);
            }
        \end{scope}

        \foreach \n in {0,1,M}{%
            \path (\n) edge[loop above] node[above=1mm] {%
                $\{ \rest \}$,\\ $\rest = \n$
            } (\n);
        }
    \end{scope}

    \node at (1.5*\gap, 0) {$\dots$};
\end{tikzpicture}
\vspace{3mm}
\caption{A sketch of the CA $A_\mathsf{battery}$, keeping track of energy.}\label{fig:ca-battery}
\end{subfigure}
\begin{subfigure}[b]{0.44\textwidth}
\centering
\begin{tikzpicture}[scale=\scale, transform shape]
    \begin{scope}[%
        every node/.style={draw,circle,minimum size=\nodesize}%
    ]
        \node (e) at (0, 0) {$\epsilon$};
    \end{scope}

    \node[left=5mm of e] (s) {};
    \path[->] (s) edge (e);

    \begin{scope}[%
        every path/.style={->,looseness=0.5,align=left}
    ]
        \path (e) edge[loop right] node[right] {%
            $\{ \mforward, \discharge \}, \top$\\
            $\{ \mbackward, \discharge \}, \top$\\
            $\{ \mright, \discharge \}, \top$\\
            $\{ \mleft, \discharge \}, \top$\\
            $\{ \mright, \mforward, \discharge \}, \top$\\
            $\{ \mright, \mbackward, \discharge \}, \top$\\
            $\{ \mleft, \mforward, \discharge \}, \top$\\
            $\{ \mleft, \mbackward, \discharge \}, \top$\\
            $\{ \turn, \discharge \}, \top$
        } (e);
    \end{scope}
\end{tikzpicture}
\caption{The CA $A_\mathsf{usage}$, modelling energy consumption}\label{fig:ca-usage}
\end{subfigure}
\caption{CAs pertaining to the battery level.}
\end{figure}

We now consider that the agent may be supplied with a finite amount of energy units $M$, in what we will for the purpose of this discussion refer to as a battery. The agent can also recharge at a designated position in the environment, but has to remain stationary to do so. To model the amount of energy left, we use the CA depicted in Figure~\ref{fig:ca-battery}. The ports $\charge$ and $\discharge$ allow manipulation of the energy state and present the new energy level as their values. To attach an energy cost to actions from the model, we need to connect this CA to other ports; the CA that takes care of this is found in Figure~\ref{fig:ca-usage}. For the sake of simplicity, we assume a unit energy cost for all actions that require energy.

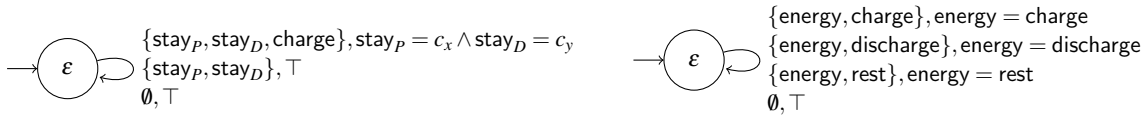
\begin{figure}
\begin{subfigure}[b]{0.49\textwidth}
\centering
\begin{tikzpicture}[scale=\scale,transform shape]
    \begin{scope}[%
        every node/.style={draw,circle,minimum size=\nodesize}%
    ]
        \node (n) at (0, 0) {$\epsilon$};
    \end{scope}

    \node[left=5mm of e] (s) {};
    \path[->] (s) edge (e);

    \begin{scope}[%
        every path/.style={->,align=left}
    ]
        \path (n) edge[loop right] node[right] {%
            $\{\mstay_P, \mstay_D, \charge\}, \mstay_P = c_x \wedge \mstay_D = c_y$\\
            $\{\mstay_P, \mstay_D\}, \top$\\
            $\emptyset, \top$
        } (n);
    \end{scope}
\end{tikzpicture}
\vspace{-4mm}
\caption{The CA $A_\charge$, modelling the charging station.}\label{fig:ca-charge}
\end{subfigure}
\begin{subfigure}[b]{0.49\textwidth}
\centering
\begin{tikzpicture}[scale=\scale,transform shape]
    \begin{scope}[%
        every node/.style={draw,circle,minimum size=\nodesize}%
    ]
        \node (e) at (0,0) {$\epsilon$};
    \end{scope}

    \node[left=5mm of e] (s) {};
    \path[->] (s) edge (e);

    \begin{scope}[%
        every path/.style={->},%
        every node/.style={align=left}%
    ]
        \path (e) edge[loop right] node[right] {%
            $\{\energy, \charge\}, \energy = \charge$\\
            $\{\energy, \discharge\}, \energy = \discharge$\\
            $\{\energy, \rest\}, \energy = \rest$\\
            $\emptyset, \top$
        } (e);
    \end{scope}\end{tikzpicture}
    \caption{The CA $A_\mathsf{energy}$, which provides the energy level.}\label{fig:ca-energy}
\end{subfigure}
\caption{More energy-related CAs.}
\end{figure}

We assume that the charging station is located at $c_y$ unit steps to the right of path position $c_x$. To model this placement, we use a CA similar to the one from Figure~\ref{fig:ca-turn-p}, found in Figure~\ref{fig:ca-charge}. In order to act on the change in energy conveniently, we use the CA in Figure~\ref{fig:ca-energy}, which fires the port $\mathsf{energy}$ whenever a port of $A_\mathsf{battery}$ fires, and provides the energy level at that port.

At this point, we can incorporate the CAs described in this section into the description of the full system. By doing so, we would obtain a system that keeps track of its battery level. If the battery is empty (state $t_1$ of $A_\mathsf{battery}$), no further energy-consuming moves are possible (for lack of a transition that fires $\discharge$). However, in the absence of a mechanism that ``plans'' the moves of the agent, such a system would most likely become stuck after it has depleted its energy. The reason for this is that there is no preference expressing that the agent likes to recharge when its energy level drops below a certain level $\ell$.

\begin{figure}[h!]
\centering
\begin{tikzpicture}[scale=\scale,transform shape]
    \begin{scope}[%
        every node/.style={draw,circle,minimum size=\nodesize}%
    ]
        \node (e) at (0,0) {$\epsilon$};
    \end{scope}

    \node[left=5mm of e] (s) {};
    \path[->] (s) edge (e);

    \begin{scope}[%
        every path/.style={->},%
        every node/.style={align=left}%
    ]
        \path (e) edge[loop right] node[right] {%
            $\{\energy\}, \energy \geq \ell, \angl{\boldzero_{E_\mathsf{return}}, \boldone_{E_\mathsf{patrol}}}$\\
            $\{\energy\}, \energy < \ell, \angl{\boldone_{E_\mathsf{return}}, \boldzero_{E_\mathsf{patrol}}}$\\
            $\emptyset, \top, \boldone_{E_\mathsf{select}}$
        } (e);
    \end{scope}
\end{tikzpicture}
\caption{The SCA $A_\mathsf{select}$, selecting between the preference regimes based on the current energy level.}\label{fig:sca-select}
\end{figure}
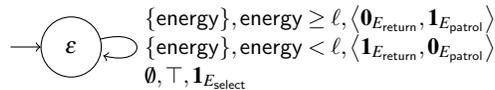

To accommodate for this, we first need to design a different regime of preferences in which the agent prefers to move towards the charging station. This can be done using a pair of SCAs, both similar to $A_\mathsf{center}$ (refer to Figure~\ref{fig:sca-center}), one to express preferences about movement along the path and another to express preferences about movement deviating from the path. For the sake of brevity, we assume the construction of this SCA to be obvious and refer to it as $A_\mathsf{return}$. We can now calculate the join composition
$$A_\mathsf{position} = (A_\mathsf{return} \otimes A_\mathsf{battery} \otimes A_\mathsf{usage} \otimes A_\mathsf{charge} \otimes A_\mathsf{energy}) \odot (A_\mathsf{move} \triangleright A_\mathsf{deviate})$$
to obtain an automaton that expresses preferences on both the regular movement and the movement toward the charging station. Finally, we must express that the preference originating from $A_\mathsf{return}$ should be considered whenever the target state of the transition within $A_\mathsf{battery}$ is $t_i$ with $i < \ell$, while the preference from $A_\mathsf{move} \triangleright A_\mathsf{deviate}$ takes precedence in all other cases. Let $E_\mathsf{return} \odot E_\mathsf{patrol}$ be the underlying semiring of $A_\mathsf{position}$, and let $h$ be the injection from $E_\mathsf{return} \odot E_\mathsf{patrol}$ into $E_\mathsf{return} \times E_\mathsf{patrol}$. Then the SCA $A_\mathsf{select}$ in Figure~\ref{fig:sca-select}, with $E_\mathsf{return} \times E_\mathsf{patrol}$ as the underlying semiring, selects a preference between the two under product composition with $h(A_\mathsf{position})$, i.e., $A_{\mathsf{agent}} = A_\mathsf{select} \otimes h(A_\mathsf{position})$.

By composing the preference value in the first position with $\boldzero_{E_\mathsf{return}}$ and the one in the second position with $E_\mathsf{patrol}$ when $\energy \geq \ell$, $A_\mathsf{select}$ essentially silences the preferences expressed by $A_\mathsf{return}$. As a consequence, the only preferences that matter when $\energy \geq \ell$ are those expressed by $A_\mathsf{move} \triangleright A_\mathsf{deviate}$.

\subsection{Discussion}

Needless to say, the final automaton can become rather large. To be precise, the automaton $A_\mathsf{agent}$ has $54K{(2L+1)}(M+1)$ states. However, in spite of syntactically available transitions, the pertinent constraints imply that not all of those states are actually reachable; for instance, positions that cannot be reached using the amount of energy $M$ available to the agent do not appear in the reachable state space. A sketch of the state space for small parameters appears in the technical report~\cite{techreport}. In this sketch, we can see that even for these small parameters the size of the state space can become quite large, with a dense transition structure. This further drives home our point that the addition of preferences is only useful when we consider preferences as first-class citizens of a compositional framework; adding preference after the fact (i.e., to the composed state space) is simply not feasible.

The vastness of this state space also indicates the difficulty of the task that an engineer faces to produce a monolithic specification of the behavior of such a system, in terms of an automaton, an LTS, a set of constraints, logic, or any other formalism.  Our compositional framework based on SCAs supports separate specification of multitudes of components, agents, aspects, features, and modules that comprise such a system --- a much more manageable task.   The specification of the whole system can subsequently be constructed using the composition operators in our framework to combine those smaller specifications. Of course, for this approach to work, one needs a verification method that is compatible with the composition operators.

\section{Conclusion}
\label{sec:conclusion}


In this paper, we argued that components with preference values associated with their actions generally aid in making a system robust, by providing alternatives when the most preferred option is not available. This property extends to composition, in the sense that such preference-aware components are capable of falling back to lower-preference actions when the composed preference dominates that of actions otherwise preferred by the component. We considered an existing formalism for preference-aware components, namely Soft Constraint Automata, and proposed a number of operators that can be used to compose such preference-aware components; each of these has its use depending on the agent being described.

By means of an example, we then illustrated the utility of SCAs and their different composition operators to design a reasonably robust autonomous agent. Most importantly, in this example we demonstrated that the use of SCAs can promote separation of concerns and extensibility. In the process, we highlighted some techniques using the composition operators, arguing in favor of their generality.

\section{Further work}
\label{sec:further-work}


For our framework to have any practical value, we of course need to be able to simulate the behavior of SCAs. This is slightly more complicated than simulating CAs, because instead of CSPs we now need to solve SCSPs to discover which transitions are available. Luckily, a number of SCSP-solving algorithms exist~\cite{bistarelli-2004}. We have developed a tentative implementation of the ideas expressed in this paper, inspired in part by the SCSP-solving approach in~\cite{sachenbacher-williams-2005}, using Gecode\footnote{See \url{http://www.gecode.org}} as a CSP-solver. We plan to continue work on this implementation, as a tool to explore and demonstrate the specification capabilities of SCAs. In particular, our tool can still be extended to include the techniques from~\cite{jongmans-kappe-arbab-2015}, which enable state-by-state evaluation of composition operators, saving computation of the unreachable state space.

We limited the example in Section~\ref{sec:example} to a single-agent system. Nevertheless, it seems intuitively clear that the same formalism can be used to describe multi-agent systems. We also suspect that in the context of multi-agent systems, coordination between all agents for every transition (to find out the most-preferred transition) is not always necessary or even possible. For instance, when the inter-agent distance is large, it may occur that transitions available to agents are never mutually exclusive and a local calculation of a most-preferred transition would suffice. Because these situations need not persist throughout the operation, it seems that some sort of mechanism to distinguish between these situations is necessary, like the minimum contact distance in~\cite{talcott-arbab-yadav-2015}.

While this work concentrates on a method to specify Cyber-Physical Systems, our framework as of yet lacks a method allowing users to verify and reason about properties of the system. To obtain such a method, one can look into existing verification techniques for CAs~\cite{baier-blechmann-klein-kluppelholz-leister-2010}. Since SCAs constitute a generalization of CAs, we expect that model checking SCAs is at least as hard as model checking CAs. A method based on rewriting logic may also be feasible, especially because it has already been shown that solving SCSPs using rewriting logic is possible~\cite{wirsing-denker-talcott-poggio-briesemeister-2007}.

In this work, we restricted ourselves to agents that make single-step decisions about their movements. As a consequence, the agents modeled are unable to handle situations in which a transition with very low preference must be taken in order for a subsequent transition with a very high preference to become available. Further work may look into accommodating a method capable of planning a multi-step path, perhaps by means of simulation as proposed by Belzner et al.\ in~\cite{belzner-hennicker-wirsing-2015}.

\bibliographystyle{eptcs}
\bibliography{bibliography}

\end{document}